\documentclass[preprint2]{aastex}
\usepackage{graphicx}

\usepackage{dcolumn}
\usepackage{bm}
\usepackage{amsmath}%
\usepackage{amsfonts}%
\usepackage{amssymb}%
\begin{document}

\title{On the low-frequency boundary of Sun-generated 
       MHD turbulence in the slow solar wind}

\author{Bidzina M. Shergelashvili\altaffilmark{1,2} and Horst Fichtner}
\affil{Institut f\"{u}r Theoretische Physik IV: Weltraum- und
Astrophysik,
Ruhr-Universit\"{a}t Bochum, 44780 Bochum, Germany}

\altaffiltext{1}{On leave from Georgian National Astrophysical Observatory,
Ilia State University, 2a, Kazbegi ave., 0160 Tbilisi, Georgia}
\altaffiltext{2}{K.U. Leuven Campus Kortrijk, E. Sabbelaan 53, 8500 Kortrijk,
Belgium}                               
\begin{abstract}
New aspects of the slow solar wind turbulent heating and acceleration are investigated. 
A physical meaning of the lower boundary of the Alfv\'en wave turbulent spectra in the solar atmosphere
and the solar wind is studied and the significance of this natural parameter is demonstrated.
Via an analytical and quantitative treatment of the problem we show that a truncation of the wave spectra 
from the lower frequency side, which is a consequence of the solar magnetic field structure and its cyclic 
changes, results in a significant reduction of the heat production and acceleration rates. 
An appropriate analysis is presented regarding the link of the considered problem
with existing observational data and slow solar wind initiation scenarios.
\end{abstract}
\section{Introduction and the physical content of the problem}\label{secphcont}
An understanding of the heating and acceleration processes of the solar
atmosphere expanding into interplanetary space as a continuous solar wind and 
solar eruptions such as coronal mass ejections is one of main targets of solar
physics. It has been
realized that the observationally determined characteristic Reynolds numbers
provide suitable conditions to develop MHD turbulence
in the strongly magnetized solar (stellar) wind. The interplanetary space
represents a {\lq}natural laboratory{\rq} able to sustain turbulent flows and
allows one to test results obtained from basic physical
grounds. In the latter context, for instance, recently \citet{chenprl10} studied
an anisotropy of the solar wind turbulence between the ion and electron
scales. \citet{kasperprl08} provided evidence in favor of the
Alfv\'en-cyclotron dissipation heating process based on the observations of
helium and hydrogen temperatures; \citet{sahraouiprl09} reported the first
direct determination of the dissipation range of MHD turbulence in the
solar wind at the electron scales. These and many other similar
studies naturally include and are complemented by 
astrophysical consequences: \citet{marinoapjl08} gave an estimation
of the rate of turbulent energy transfer, which can contribute to the in situ heating of the wind;
\citet{luoapjl10} revealed the scale-dependent anisotropy
of the wave power spectrum. In addition, suprathermal ions and plasma
wave spectra upstream of interplanetary shocks driven by coronal mass ejection
events have been analyzed by \citet{bamertapjl08}. 

Already
in the 70s of the last century the
turbulent cascade of Alfv\'en waves has been suggested as an efficient
process to transfer energy from large-scale waves to small-scale ones,
which are more easily subject to damping by different transport
processes possibly operating in the plasma flow. Fundamental work, particularly
for the solar context, has been done
by Tu and collaborators \citep{tujgr84,tumarsch96} by
developing basic aspects of the theory of MHD
turbulence in the solar wind. More recent works demonstrate a consistency
of the wave heating process
with the observed temperature and velocity especially in coronal
holes \citep{hujgr99,vainioaa03}. We know that similar processes
can operate in stellar atmospheres
\citep{narainrev94,ulmshneiderapj01,elfimovapj04}. Therefore, proper modeling of
the turbulence is of broad astrophysical interest. 

When the modeling concerns the equatorial regions (even at solar
minimum characterized by the `double structure' of slow and fast wind) the wave heating 
scenario is rather complex because of several reasons. Firstly, at low latitudes the magnetic
field topology drastically differs from that of the polar regions and, secondly, there is clear
observational evidence for multiple sources sustaining the observed heating and
acceleration rates in equatorial regions \citep{kasperapj07} through the solar cycle. Many
analytical and numerical models use artificial ad-hoc functions
\citep[see, e.g.,][]{holstapj07,kleinmanannge09} mimicking the observed
latitudinal variation of the solar wind structure at solar minimum.
The modeling is even more complex close to solar maximum when the solar wind becomes a complicated
mixture of `channels' of fast (above coronal holes) and slow (above coronal streamers) wind regions.

It is well known that the solar (and plausibly many stellar) atmosphere(s) is
(are) abundantly populated with Alfv\'en waves 
\citep[see, e.g.,][]{Cirtain-etal-2007,dePontieu-etal-2007}.
 
A stochastic nature of the wave excitation and propagation processes naturally
led in the past to the notion of MHD turbulence in the context
of the heating and acceleration of the solar wind. It is now conventionally accepted
that an engagement of the clearly observed Alfv\'en wave spectrum into the
turbulent cascades represents one of the important processes
\citep[e.g.,][]{chenprl10,marinoapjl08,oughtmatt05} for a transmission
of the wave energy from large (non-dissipative) to small (damping) scales
\citep{kasperprl08}.

Non-modal cascading was suggested by \citet{shergself06} as an alternative process
that can be specifically relevant for the boundaries of coronal holes in the regions where
the transition from fast to slow wind occurs. Actual realizations of such processes and their proper
testing in comparison with observational data is one of the paramount challenges
of modern solar (and generally stellar) physics.

In order to perform a proper analysis of the solar wind formation processes
one should model the complex and
very dynamic structure of the solar atmosphere and outflowing wind. There are
many processes with a wide range of characteristic spatio-temporal scales which govern the complex
topological configuration of the magnetic field.
The aspects playing a dominant role
in the distribution and properties of the Alfv\'en wave sources are the time-varying streamer belt and the
related heliospheric current sheet. At solar minimum the streamer-like structures and the current sheet are 
concentrated
in the vicinity of the ecliptic plane. This configuration reproduces the double (slow-fast) structure of the
solar wind.
Namely, at low latitudes the slow wind is generated (with characteristic velocity at solar minimum of about 
400 km/s), while almost all of the rest of the solar surface in both hemispheres
is covered by huge `polar' coronal holes giving origin to the fast wind (of about 800 km/s).
There is a relatively sharp boundary between these two zones.
At this stage there are no spatial constraints on the wave sources as the central streamer structures are
rather wide and maintain a larger range of frequencies within the streamers. In general, however, 
we suppose that there is still some restriction on the frequency domain (as we model below)
within the streamer belt even at the cycle minimum manifested by the presence of the slow wind at low latitudes.
However, with the advance of the solar cycle this topological picture
changes drastically. In particular, the latitudinal width of the streamer generating zone widens gradually reaching 
very high latitudes at solar maximum (up to 70$^{\circ}$ and even further). This zone
becomes more densely populated with the streamer-like structures with
significantly smaller transverse spatial scales
compared to those at solar minimum. As a result, characteristic wavelengths (frequencies) of waves excited in these
structures can be bound from above (below). Observations clearly evidence that the shape of the solar wind
evolves accordingly \citep{richardson01}. The velocity profile becomes {\lq}homogeneous{\rq} (characteristic speed 600 km/s) 
throughout the streamer zone and the transition to the high speed wind (800 km/s) occurs only within small polar
caps to which the polar coronal holes are confined. 

The second aspect contributing to the shaping of the Alfv\'en wave spectra are active regions
hosting other open field structures, which mainly can contribute to the high
frequency part of the spectrum. The resulting magnetic
structures as sources of the waves are known as active region sources
\citep{liewersph04}. The proper and analytically consistent modeling of this factor
requires a separate effort bringing the issue beyond the scope of the current paper. Here we point out at least
some implications how the contribution from the active region sources could be modeled, and a rigorous study of
the issue will be published elswhere. In the present paper, which is based on the simple analytical
framework for the solar wind wave turbulent heating developed by \citet{vainioaa03}, 
we focus on the effect of the frequency truncation from below,
which must be expected to have an 
effect on the overall process of the wave cascading and damping when this 
lower ferequency boundary is located within the inertial range. 
The range of Alfv\'en wave frequencies observed in the solar atmosphere is determined by wave excitation and
damping. The wave damping has received great attention in the past and is well elaborated in the related literature
(including the citations above), and the analyses carried out in those works clearly
indicate the significance of the high-frequency (small spatial scale) boundary position, which is determined by the
interconnection between the spatial scales of waves and damping
processes at work. In order to make clear the subject of the analysis that we
are presenting in this paper we should state that, we focus here on the low-frequency boundary of the domain based
on physical grounds, and we treat it as
a natural parameter playing a significant role for the relative efficiency of 
turbulent processes in different regions of the solar atmosphere. The locations of
the latter depend on the transversal spatial scales of the magnetic field structures and related
wave excitation sources, which are responsible for the creation of the wave
spectrum. This issue requires more attention because its significance
usually has been diminished  in existing models. This is the main goal of the
proposed analysis, and therefore in what follows we give a mathematical formulation 
of the above statements and parametrize the modeling in terms of
the relevant physical quantities using a notation that is conventional for
models of the wave spectral power diffusion due to the turbulent cascade
in the solar atmosphere.
\section{The model equations}\label{modeq}
A quantitative model of the contribution of the MHD
turbulence into the total heat balance of the expanding solar atmosphere has been
developed
and condensed into the balance equation governing a steady-state spatial
configuration of the  spectral wave power $P$ \citep{tujgr84}: 
\begin{equation}\label{tusteady}
  {\vec \nabla} \cdot [({\vec u}+{\vec v_A}) P] + \frac{P}{2}
({\vec \nabla} \cdot {\vec u}) = - \frac{\partial F}{\partial f}
\end{equation}
where $\vec{u}$ and $\vec{v}_A$ are the background and the Alfv\'en velocity,
respectively, and $F$ represents a 'cascading' of the spectral power because
of the presence of the turbulent cascade. For an explicit form of the cascading
function $F$ see, e.g., \citet{vainioaa03}. This equation usually is
generalized
to its dynamical counterpart to obtain the temporal evolution of the spectral power
\citep{hujgr99}:
\begin{equation}\label{tudynam}
  \frac{\partial P}{\partial t} + {\vec \nabla} \cdot [({\vec u}+{\vec v_A}) P]
+ \frac{P}{2}
({\vec \nabla} \cdot {\vec u}) = - \frac{\partial F}{\partial f}
\end{equation}
Following the conventional notation we define the pressure produced by waves as:
\begin{equation}\label{wavepress}
 p_w =\frac{1}{8\pi}\int _{f_0} ^{f_H} P {\rm d}f.
\end{equation}
Here, $f_0$ and $f_H$ are the 'boundaries' of the considered wave spectrum.
These two parameters play key roles in the wave heating models and
they include links between the Alfv\'en wave spectrum frequency domain and the
physical conditions in the environment where those waves propagate and
interact. In particular, $f_0$ represents a minimal frequency at which
the wave energy is injected into the cascade and it contains information on the
characteristic spatial and temporal scales of the wave sources, while $f_H$
is a maximum frequency beyond which the turbulent spectrum is truncated
because of strong damping and it manifests properties and typical scales at
which the considered wave dissipation processes are active. We turn to this issue
again below. 

   Taking the integral of equation (\ref{tudynam}) w.r.t. wave frequencies within the
mentioned frequency range $f_0 < f < f_H$, after straightforward
manipulations and taking into account that
\begin{equation}\label{intanz1}
 \int _{f_0} ^{f_H} \vec \nabla P {\rm d}f = \vec \nabla \left ( \int _{f_0}
^{f_H} P {\rm d}f \right ) + P(f_0)\vec \nabla f_0-P(f_H)\vec \nabla f_H
\end{equation}
as well as
\begin{equation}\label{intanz2}
 \int _{f_0} ^{f_H} \frac{\partial F}{\partial f} {\rm d}f = F(f_H)-F(f_0)
\end{equation}
one arrives at the familiar equation for the wave pressure:
\begin{equation}\label{eqwp}
 \frac{\partial p_w}{\partial t} + {\vec \nabla} \cdot [({\vec u}+{\vec v_A})
p_w] + \frac{p_w}{2} ({\vec \nabla} \cdot {\vec u}) + \frac{Q_w}{2}=0
\end{equation}
with heating contribution $Q_w$ from the Alfv\'en wave spectrum, which we write
in general form as:
\begin{equation}\label{heatq}
 Q_w = Q_{w1}+Q_{w2} = \frac{1}{4\pi} \left [ F(f_H)-F(f_0)\right ]  +
\frac{({\vec u}+{\vec v_A})}{4\pi} \cdot \left [ P(f_0)\vec \nabla f_0-P(f_H)\vec \nabla f_H \right ] 
\end{equation}
In the standard models of the solar wind wave heating main
attention has traditionally been given to the spectral truncation limit $f_H$,
and
corresponding explicit expressions (derived on the ground
of the underlying microscopic transport processes in a plasma, say cyclotron
damping etc.) for this frequency have been given in the related literature
\citep[see, e.g.][]{hujgr99}.
Even a gradient of this limit along the radial distance from the Sun has been
taken into account manifesting itself in the erosion (shrinking from the high
frequency side) of the turbulent spectrum. However, a contribution from the
lower frequency limit $f_0$ has always been omitted. 
It was assumed that the characteristic frequency, where the actual wave spectrum begins, is constant --
implying a vanishing gradient of $f_0$ in the expression (\ref{heatq}) -- and at values significantly
below the inertial range of the spectrum within which the MHD turbulence
operates effectively and as a result waves mainly follow the WKB behavior. This
latter assumption removes the wave power cascading term $F(f_0)=0$ and leads,
therefore, to a simplified version
of (\ref{heatq}) as given, for instance,
in \citet{hujgr99}. In the latter work $f_0$ is even set to zero
with the corresponding physical reasoning in the fast solar wind.

We agree that, in principle, the above argumentation can be valid, but 
only for the case when the lower boundary of the spectrum always remains outside the inertial range.
However, we argue here that this assumption cannot be true for the slow solar wind regions populated with
curved and isolated islands of magnetic structures even if both terms containing
$f_0$ in expression (\ref{heatq}) remain vanishing because of no cascading 
to the cut-off frequency from lower frequencies (leading to $F(f_0)=0$ and $P(f_0)=0$).
The effect of the spectral truncation still plays a
role when $f_0$ appears within the inertial range, as it systematically enhances the wave
power deficit with increasing distance from the Sun. Many
theoretical and numerical models of the solar wind include physical quantities set
artificially as discontinuous functions of latitude for the sake of mimicking
the real structure of the solar outer atmosphere. However, if one
wants to develop a consistent model for the evolution of the solar
wind, then gradual changes 
with heliographic latitude
in the topology of the magnetic fields and corresponding scales of the Alfv\'en
wave sources should be taken into account. Therefore, our main intention
is to study wave turbulent processes in connection with the variable source properties and probabilities
of their appearance during the solar cycle. In this study we use simple modeling of the mentioned variability, in order to
create a framework for further more complex modeling of the dynamical processes in the solar atmosphere
responsible for the formation of the solar wind. Understanding both the methodology
how to do this and possible ways of realizations linked with the actual
spatial distribution of the observable physical quantities represent the 
goals of the present paper.

\begin{figure*}
  \includegraphics[scale=0.8]{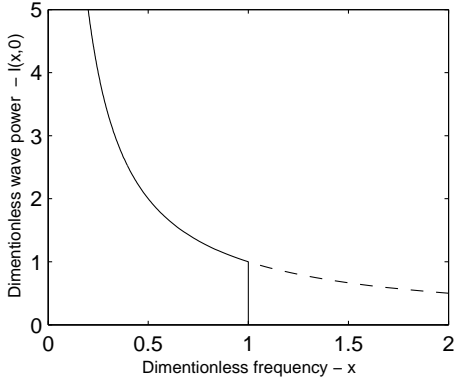}\\
  \caption{Plot of the function $g(x)=I(x,0)=H(1-x)/x$ (solid line). The dashed line shows the part of the spectrum
  that has been truncated.}\label{fig_epsalspe}
\end{figure*}
\section{Solution of the wave transport equation}\label{sectupdsol}
\subsection{A class of solutions with a low-frequency cut-off}\label{sunsectsolclass}
For the subsequent analysis we require a solution of the wave transport equation that 
includes a lower frequency boundary of the wave power spectrum. Like in the reference model 
\citep{vainioaa03} we start from the equation:
\begin{equation}\label{wavetransp}
 \frac{\partial I}{\partial \tau} - I^{1/2}\frac{\partial I}{\partial x}=0,
\end{equation} 
where, the function $I$ and the independent variables $\tau$ and $x$ are defined like in \citet{vainioaa03}, i.e.:
\begin{equation}\label{dimensionlessx}
x = \left(\frac{f_0}{f}\right)^{2/3} 
\end{equation}
\begin{equation}\label{dimensionlesstau}
\tau = 2\pi f_0 \epsilon_p^{1/2} \int\limits_{r_{\odot}}^r
                       \frac{C^2(r^{\prime}) V_{\odot} v_A(r^{\prime})}{V^3(r^{\prime})}
		       \left(\frac{n_{e\odot}}{n_e(r^{\prime})} \right)^{1/4} dr^{\prime} .
\end{equation} 
where we have chosen the normalisation frequency as $f_n = f_0$.
The above equation (\ref{wavetransp}) describes a simple `dynamics' of the dimensionless wave power $I$ in the
($x,\tau$)-space and can be solved by the method of characteristics, implying the
following set of equations:
\begin{equation}\label{charact1}
dI = 0, 
\end{equation} 
\begin{equation}\label{charact2}
d\tau = dx/I^{1/2}. 
\end{equation} 
As in the reference model, this implies that the
characteristic ``time'' variable (we refer to this variable as time only formally, it actually corresponds to 
the heliocentric distance $r$) satisfies $\tau '=\tau$ and that the ``Lagrangian'' differential of $dI=0$ vanishes
so that $I=I_c={\rm const.}$ as well as $x+I^{1/2}\tau =x_c={\rm const.}$ along stream lines in the ($x,\tau$)-space.

Therefore, to find the solution one should look for it in the form:
\begin{equation}\label{gfunc}
 I=g(x+I^{1/2}\tau),
\end{equation}
where
\begin{equation}\label{ginit}
 g(x)=I(x,0)
\end{equation}
determines the shape of the spectrum at the solar surface $\tau=0$.
For the sake of clarity we invert the analysis and search the solution $x(I,\tau)$
of Eq.\ref{charact2}, which is:
\begin{equation}\label{xcharactsol}
 x(I,\tau)=F(I)-I^{1/2}\tau,
\end{equation}
with $F(I)$ being the inverse function of $g(x)$ and $\tau< I^{-1/2}F(I)$ determining a characteristic distance
of the wave transport, where the waves reach certain small scales $x\rightarrow 0$ and dissipate.
In order to make our current derivation directly comparable with the reference model we
define the spectrum at the surface as
\begin{equation}\label{gfuncgeneral}
 g(x)=I(x,0)=H\left(  1-x\right)  x^{\frac{3q-5}{2}}=H\left(  1-x\right)g_0(x),
\end{equation}
where $H$ denotes the Heaviside function, $q < 5/3$ \citep{vainioaa03}, the subscript 0 indicates functions defined as
in the reference model, and
the value $x=1$ corresponds to the frequency $f_0$ where the spectrum is cut abruptly
because of the deficit of low-frequency waves in the spectra, incorporated by the factor $H(1-x)$. The function
$g(x)$, which is shown for $q=1$ in Figure~\ref{fig_epsalspe}, corresponds to the following spectrum at the surface:
\begin{equation}\label{spectrumin}
P(f,r_{\sun})=H\left(  1-\left(
\frac{f_{0}}{f}\right)  ^{\frac{2}{3}}\right)  \frac{\epsilon_{p} B_{\odot}^{2}}{f_{o}}\left(  \frac{f_{0}}{f}\right)
^{q}
=H\left(  1-\left(
\frac{f_{0}}{f}\right)  ^{\frac{2}{3}}\right) P_0 (f,r_{\sun}).
\end{equation} 

The function $F(I)$ can be represented in this case in the following form:
\begin{equation}\label{Ffunc}
F(I)=\left\{
\begin{array}
[c]{ll}%
1 & \text{if }0<I<1\\
I^{\frac{2}{3q-5}} & \text{if }I\geq1
\end{array}
\right. 
\end{equation}
Using the solution (\ref{xcharactsol}) we arrive at a set of algebraic equations
governing the relation between $I$, $x$ and $\tau$ within the entire domain under consideration:
\begin{equation}\label{eqgeneral}
 \left\{
\begin{array}
[c]{ll}%
1-\left[  x\left(  I,\tau\right)  +I^{1/2}\tau\right]  =0, \hskip0.3cm \tau< I^{-1/2} & \text{if }0<I<1\\
I^{\frac{2}{3q-5}}-\left[  x\left(  I,\tau\right)  +I^{1/2}\tau\right]  =0, \hskip0.3cm \tau< I^{\frac{3(3-q)}{2(3q-5)}} &
\text{if }I\geq1
\end{array}
\right.  
\end{equation} 
It is straightforward to show that for $x\gg I^{1/2}\tau$ the approximate solution
is $I(x,\tau)\approx I(x,0)$ and that we obtain the same WKB behaviour as in the reference model.
The only difference is that the spectrum $P(f,r_{\sun})$ is truncated from below at the frequency $f=f_0$.

Now we turn to the other limit $x\ll I^{1/2}\tau$. In this case one finds:
\begin{equation}\label{solgeneral}
I\approx \left\{
\begin{array}
[c]{ll}%
\frac{1}{\tau^{2}} & \text{if }\tau>1\\
\tau^{\frac{2\left(  3q-5\right)  }{3\left(  3-q\right)  }} & \text{if
}\tau\leq1
\end{array}
\right.  
\end{equation} 
which can be readily rewritten as:
\begin{equation}\label{solspect}
P\left(  f,r\right) \approx \left\{
\begin{array}
[c]{ll}%
  P_{0}\left(  f,r\right) \left(  \frac{f_{g}}{f_{0}}\right)  ^{\frac{4}{3}}=
P_{\rm{WKB}}(f,r) \left(  \frac{f_{g}}{f_{0}}\right)  ^{3-q}\left(
\frac{f_{0}}{f}\right)  ^{\frac{5}{3}-q} & \text{if }\tau>1\\
P_{0}\left(  f,r\right)  =P_{\rm{WKB}}(f,r) \left(  \frac{f_{g}}{f}\right)  ^{\frac{5}{3}-q} & \text{if
}\tau\leq1
\end{array}
\right.,   
\end{equation} 
where $f_{g}=f_{0}\tau^{\frac{2}{q-3}}$ is a generalized breakpoint frequency. 
\subsection{Comparison with the reference model}\label{sunsectsolpart}
In the previous subsection we have derived a class of solutions of the wave transport equation, parametrized by the value of
the exponent $q$, taking into account a truncation of the spectrum towards low frequencies. In order to 
perform a direct comparison with the quantitative analysis carried out in the reference model, we briefly list the relevant 
equations for the solution for $q=1$ (see Figure~\ref{fig_epsalspe}): 
\begin{equation}\label{gpart}
 g(x)=\frac{H\left(  1-x\right)  }{x}
\end{equation} 
\begin{equation}\label{Ppart}
P(f,r_{\odot})=H\left(  1-\left(  \frac{f_{0}}{f}\right)
^{\frac{2}{3}}\right)  \epsilon_{p}\frac{B_{\odot}^{2}}{f}   
\end{equation} 
\begin{equation}\label{Fpart}
 F(I)=\left\{
\begin{array}
[c]{ll}%
1 & \text{if }0<I<1\\
\frac{1}{I} & \text{if }I\geq1
\end{array}
\right. 
\end{equation} 
In the limit $x\ll I^{1/2}\tau$ one has:
\begin{equation}\label{solpart}
I\approx \left\{
\begin{array}
[c]{ll}%
\frac{1}{\tau^{2}} & \text{if }\tau>1\\
\frac{1}{\tau^{\frac{2}{3}}} & \text{if }\tau\leq1
\end{array}
\right.  
\end{equation} 
and 
\begin{equation}\label{solpartspectr}
P\left(  f,r\right)  \approx \left\{
\begin{array}
[c]{ll}%
P_{0}\left(  f,r\right)  \left(  \frac{f_{c}}{f_{0}}\right)  ^{\frac{4}{3}}=
P_{WKB}\left(  \frac{f_{c}}{f_{0}%
}\right)  ^{2}\left(  \frac{f_{0}}{f}\right)  ^{\frac{2}{3}} & \text{if }%
\tau>1\\
P_{0}\left(  f,r\right)  =P_{WKB}\left(  \frac{f_{c}}
{f}\right)  ^{\frac{2}{3}} & \text{if }\tau\leq1
\end{array}
\right. 
\end{equation} 
where $f_c=f_0/\tau$ is the breakpoint frequency as defined in the reference model.

There are several conclusions one can draw based on the above derivation.
In particular, the solutions given in the reference model are independent of the frequency normalization
constant $f_n$, while the new solution explicitly depends on it. 
As the most convenient value we have chosen
$f_n=f_0$. Further, we can see that the truncation of the initial spectrum results in a significant
modification of the spectra at all heliocentric distances and, consequently, one can expect a related 
modification of the heating as well as acceleration rates.
Exactly the quantitative investigation of the latter is the subject of the following sections.
\section{A variable lower boundary of the frequency domain}\label{sectfreqdom}

We start our analysis from the concept of structured magnetic fields
within the streamer zone as opposed to the non-structured
{\lq}straight{\rq}
magnetic
fields in coronal hole regions. It is apparent that the characteristic rate of
curvature of the magnetic field should depend on the statistical distribution
of the magnetic structures fed by means of two main known contributors -
shearing
due to the random velocity fields and random emergence of the magnetic flux.
The topological structure of the magnetically governed zones, of course, also
depends on the phase of the solar cycle which leads to the time variable
strength of the magnetic field and to a corresponding variation of the internal 
structure of the streamer structures. 

  As we have mentioned above, the presence of the lower frequency boundary 
in the spectra contains information on the scales and distribution of wave sources 
across the solar disk. This information can be imported into the model from the
observations of the distribution of magnetic structures apart from
the turbulent cascade process itself. In order to
calculate properly the lower frequency limit we have to establish a relation between the
location of the magnetic structures developed during the solar cycle and the scales of wave sources.
    
Before we calculate the physical quantities determining characteristic temporal
and spatial scales of wave sources it has to be mentioned that we use \citet{vainioaa03} 
as a reference model and throughout the following text
we adopt evaluations of physical quantities taken from the subsection 2.1 of that work. 

\begin{figure*}
  \includegraphics[scale=0.9]{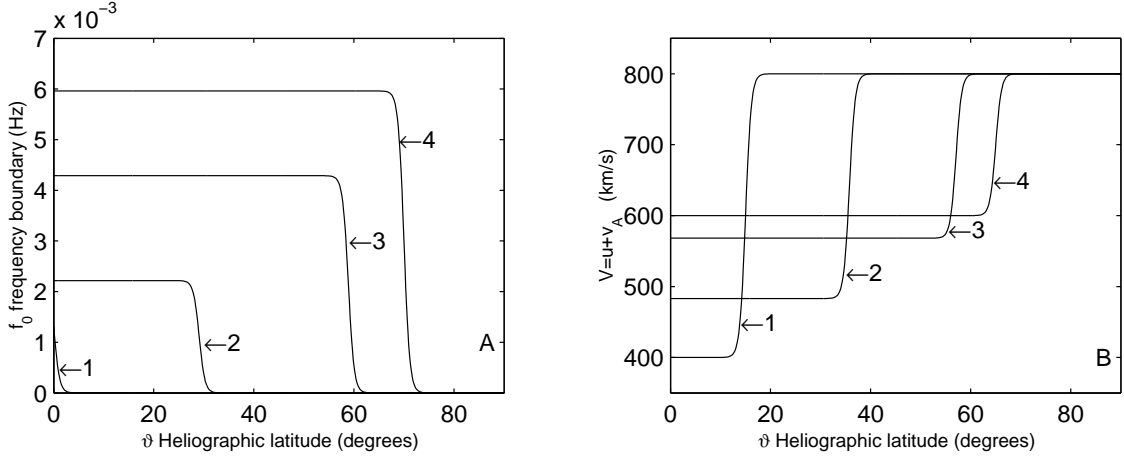}\\
  \caption{Panel A: Sketch of the latitudinal distribution of the lower boundary of the frequency spectrum.
           The curves correspond to four different phases of the solar cycle: $t^{\prime}=0, 1.5, 3.5$ and 
           $5.5$ (years counted from minimum activity) labeled 1 to 4, respectively. Panel B: Sketch of the observed distribution of
           the total speed $V$ vs.\ latitude. The four curves are shown in the same order as in Panel A.}\label{fig_freqvel}
\end{figure*}

We formulate a simple model of the observed variability of several physical quantities
assuming the 
corresponding distribution to be related to the wave source probability distribution over latitude
within both streamer zones and coronal holes. For the sake of simplicity we use the
smoothed Heaviside function:
\begin{equation}\label{heaviside}
 H(x)= \frac{1}{1+e^{-2kx}}, x\in(-\infty,\infty)
\end{equation} 
where $k$ is the parameter determining the sharpness of the transition. 
The dynamical changes in the magnetic field topology (as described in the introduction)
should impose constraints on the available scales $Sc$ (measured in Mm) of the wave sources, which we
approximate in the following manner:
\begin{equation}\label{wscale}
 Sc=\frac{300-200\vert \sin(\frac{\pi}{11}t^{\prime}) \vert}{H(\vartheta _2-\vartheta)},
\end{equation}
here $\vartheta _2= (7\pi/18)\vert \sin(\pi t^{\prime}/11) \vert$ is the location of
the transition from the zone of
finite $f_0$ to that of a vanishing one. At the solar minimum ($t^{\prime}=0$) $\vartheta _2 =0$ represents the fact that
there is no truncation of the spectrum even if streamers are at the equator. We
calculate the lower frequency boundary
of the spectrum (using the Alfv\'en speed) as $f_0=v_A (r_{\sun})/Sc$ to obtain four curves of $f_0$ vs.\ latitude, each corresponding to
four different stages of the solar cycle:
$t^{\prime}=0$, $1.5$, $3.5$ and $5.5$ years. We plot the corresponding curves of $f_0$ in Figure~\ref{fig_freqvel}, Panel A,
for the mentioned stages of the solar cycle.

Further we write $V = u + v_A$ as a function of time:
\begin{equation}\label{vtime}
 V(t^{\prime},\vartheta)=V_1(t^{\prime}) + (800-V_1(t^{\prime})) H (\vartheta - \vartheta_1),
\end{equation}
where $\vartheta _1 = \pi /12 +(5\pi/18)\vert \sin(\pi t^{\prime}/11) \vert$ is the
variable boundary of the velocity transition,
$t^{\prime}$ is time measured in years, $k=50$, $V$ and $V_1$ are measured in km/s and
\begin{equation}\label{v1}
 V_1(t^{\prime})= 400 + 200 \left \vert \sin \left ( \frac{\pi}{11}t^{\prime}\right ) \right
\vert.
\end{equation}
Corresponding curves are shown in Figure~\ref{fig_freqvel}, Panel B, in the same order
as described above.

In order to demonstrate how the above scenario operates in practice and to prove the validity
of the approach we perform further a quantitative analysis based on the
proposed updated model of the solar wind turbulent heating.

\begin{figure*}
  \includegraphics[scale=0.9]{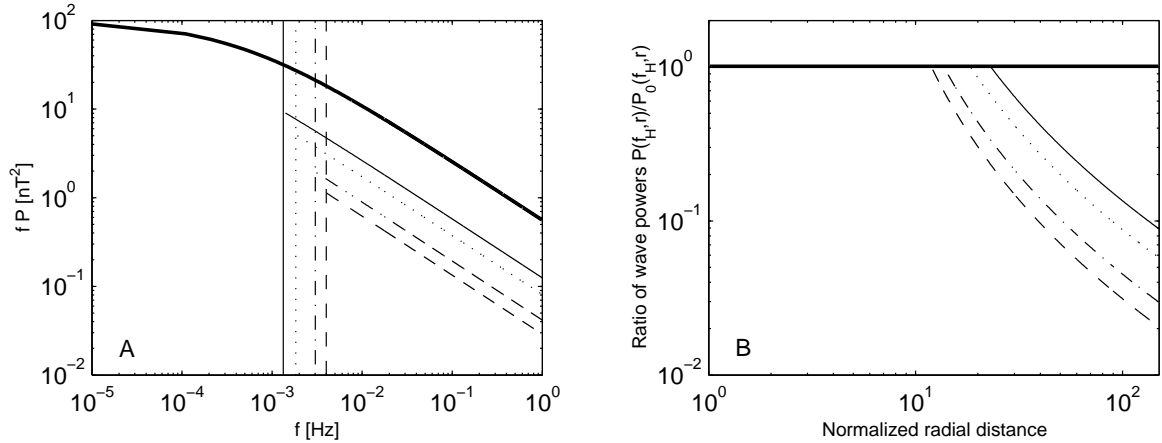}\\
  \caption{The linestyles correspond to four different stages of the solar cycle in the following order: 
           $t^{\prime}=0$ (thin solid line), $t ^{\prime}=1.5$ (dotted line), $t^{\prime}=3.5$ (dashed-dotted line) and
           $t^{\prime}=5.5$ (dashed line). Panel A: The scaled wave power spectra ($f P$) vs.\ frequency at 0.3~AU. The thick solid line is 
           the complete spectrum of the reference model. The vertical lines mark the cut-off frequencies $f_{0}$ at four different stages
           of the solar cycle. Panel B: Ratio of the spectral powers, at the corresponding dissipation frequencies,
           calculated using the considered frequency domain truncation and the
           one taken from the reference model:  The difference between the reference values (thick horizontal line ) and
           the updated turbulence spectra (the curves) is a result of their low-frequency truncation within the inertial range.  
           }\label{fig_spectr}
\end{figure*}

\section{Quantitative analysis and proof of the concept's validity}\label{sectquantproof}
 
We aim to perform an appropriate quantitative study of the problem 
in order to obtain a consistent proof of the proposed concept. 
As we study the possibility of the presence of lower frequency cut-offs in the Alfv\'en wave spectra
above streamer-like structures, it is natural to assume that the locations of such restricted sources are
linked with the latitudinal extent of the streamer generating zone and its variation with the solar cycle.

In this paper we show how the frequency cut-off affects the wave spectrum as well as the 
heating and acceleration rates. Therefore, like in \citet{vainioaa03}, we use $V=400$ km s$^{-1}$ as a target velocity.
Using this value we  derive four characteristic values of the cut-off frequencies: 1) $f_0=10^{-6}$ Hz. This is the case
recovering the one in reference model (case of no truncation) as in this case the cut-off frequency remains outside the inertial range along the entire
domain of computation so that there is no modification of the spectrum. 2) the other valuas are calculated 
and shown in Panel A of Fig.~\ref{fig_spectr} as vertical lines at $f_0=1.3\cdot 10^{-3}$ Hz. (solar minimum, $t^{\prime}=0$ (solid line),
$f_0=1.8\cdot 10^{-3}$ Hz ($t ^{\prime}=1.5$ (dotted line)), $f_0=3.0\cdot 10^{-3}$ Hz. ($t^{\prime}=3.5$ (dashed-dotted line)) and
$f_0=4.0\cdot 10^{-3}$ Hz. (solar maximum, $t^{\prime}=5.5$ (dashed line))). In the same panel we plot the spectra at
0.3~AU corresponding to these five cases. In panel B of Fig.~\ref{fig_spectr} we plot the ratio of the spectral powers
calculated using the considered frequency domain truncation and the
one taken from the reference model. The points of break at which the curves corresponding to the updated spectra start
to deviate from the reference values (the thick horizontal line) mart the distances at which $\tau = 1$
($f_c= f_0$). These curves were obtained using updated spectra acurately calculated in this paper
and given by expression (\ref{solpartspectr}) (case $q=1$). 
In \citet{vainioaa03} it has been shown that 
there is only a small difference between convective and diffusive formulations of the flux function $F$.
We concentrate on the first approach and consequently use:
\begin{equation}\label{flux}
F=2\pi C^2 \frac{v_A}{V}\frac{f^{5/2} P^{3/2}}{B}.
\end{equation} 
The physical quantities in the latter expression are defined as follows:
\begin{equation}\label{power}
P\left(  f,r\right)  \approx \left\{
\begin{array}
[c]{ll}%
\frac{P_{\rm{WKB}}(f,r)}{1+\left [ f/f_c (r)\right ]^{2/3}} & \text{if }f_0\leq f_c\\
\frac{P_{\rm{WKB}}(f,r)}{1+\left [ f_0/f_c (r)\right ]^{2}\left [ f/f_0 \right ]^{5/3}} & \text{if }%
f_0> f_c
\end{array}
\right. 
\end{equation}
with
\begin{equation}\label{pwkb}
 P_{\rm{WKB}}(f,r)=P(f,r_{\sun}) \frac{B(r)v_A (r)}{B_{\sun}v_{A\sun}},
\end{equation} 
representing the part of the spectral power
outside the inertial range where waves follow predominantly a WKB behaviour and where
\begin{equation}\label{psun}
P(f,r_{\sun}) = \varepsilon _P \frac{B_{\sun} ^2}{f} H\left(  1-\left(
\frac{f_{0}}{f}\right)  ^{\frac{2}{3}}\right),
\end{equation} 
and $f_c (r)$ is the breakpoint frequency 
at which the Kolmogorov type of the spectrum starts to prevail:
\begin{equation}\label{fc}
 \frac{1}{f_c (r)}=2\pi \varepsilon _P ^{1/2} \int _{r_{\sun}} ^r \frac{C^2 (r') v_A(r')}{V^2} 
\left ( \frac{n_{e\sun}}{n_e (r')}\right ) ^{1/4} dr' .
\end{equation} 
Here $\varepsilon _P$ is a dimensionless spectral parameter determining the intensity of wave excitation. 
Its value in the case of absence of additional wave sources is set to $5 \cdot 10^{-5}$. 
In general, a more complete model of the solar wind
must include also angular components of velocities, however for the sake of simplicity
we assume those components to be vanishing. 
$C^2 (r) = \alpha \alpha _1 (r)$ is a model parameter, where $\alpha$ is a cascading constant and
\begin{equation}\label{alph1}
 \alpha _1 = \left\{
  \begin{array}{l l}
    \alpha _0 \left( \frac{r-r_{\sun}}{9r_{\sun}}\right) & \hskip0.2cm r_{\sun} \leq r < 10r_{\sun}\\
   \alpha _0 & \hskip0.2cm r \geq 10r_{\sun}\\ 
  \end{array} \right.,
\end{equation}
with $\alpha _0$ another dimensionless parameter determining the ratio of outward and
inward propagating wave intensities. Its value when additional sources are absent is set to 0.05.

We have also performed computations for the solar wind acceleration due to the wave pressure gradient ${\vec \nabla} p_w$.
As in \citet{vainioaa03}, we use explicit the expression:
\begin{equation}\label{pw}
p_w=\frac{v_A}{v_{A\sun}} \frac{A_{\sun}}{A} p_{w\sun} - \frac{v_A}{2VA} \int _{r_{\sun}} ^r 
\frac{A(r')}{v_A(r')}Q(r')dr',
\end{equation}
where $A\propto 1/B$ is a cross sectional area of the open field structure.
The results of calculations are shown in Fig.~\ref{fig_Q_new} Panel B. Again, 
the thick solid curve shows the radial distribution of the wave pressure gradient
$-(m_p n_e)^{-1} \partial p_{w0} / \partial r$ corresponding to the case of no cut-off
and, thus, the reference model.
The calculations for $-(m_p n_e)^{-1} \partial p_{w} / \partial r$  
are done for the above-mentioned four epochs of the solar cycle
shown by the thin curves
with the same order of the linestyles as in Panel A. 
It is seen that the acceleration rate decreases significantly
(according to these calculation by 15-25 percent) within the first ten solar radii from the solar surface, 
while it monotonically decreases in absolute value beyond this distance. To summarise this part of our
investigation we conclude that the truncation of the available wave spectrum from the lower frequency side
leads to a significant decrease in both heating and acceleration rates and, therefore, this proves the 
main message of this paper that the effect of the lower frequency boundary $f_0$ should not be abandoned in
the slow wind models. 

The analysis
above lets us conclude that if only one kind of wave sources would operate in the
equatorial zone then the appearance of more and more streamer-like open structures with increasing solar
activity -- leading to a gradual decrease of their transversal spatial scales -- would result in significant cooling
of the wind plasma above the distance of 10 solar radii and deceleration of the wind stream even within that zone
(what is actually observed outside the sunspot generating zone at high latitudes \citep{richardson01}). 
However, observations
evidence  \citep{kasperapj07} that there clearly operate two kinds of wave sources with relative significance at different stages of
the solar cycle. It is reasonable to assume that these growing number of small-scale magnetic structures in the vicinity
of active zones increases the probability of reconnections, on the one hand, and
leads to creation of additional small-scale open field structures anchored in active zones, on the other hand. It is conventionally
presumed that those reconnection processes shake these open structures and, thus, represent so-called active region
sources of additional wave power. A significant number of such small structures has a reasonable potential
to compensate the losses of energy arising from the truncation of the lower frequency part of the spectrum.
In other words, even though for low frequency modes the equatorial part of the atmosphere is not accessible,
increased wave power is nevertheless available in the high-frequency part of the spectrum above the cut-off
frequencies $f_0$. One could ask here: what should be an actual value of the higher frequency wave power
to compensate the losses in thermal and mechanical energy production rates obtained above? As we mentioned above
this issue stays outside the scope of the current study.
It should be noted that
observed wave spectra \citep[e.g. see in][]{tumarsch96} contain a combined wave power supplied by
the sources from entire coronal holes (including polar regions) and some possible local sources not located
on the Sun, but in the solar wind. This is why, possibly, the truncation of the power spectra as shown in Figure~\ref{fig_spectr}
is not observationally resolvable so far. However, large error bars in the spectra below the frequencies 10$^{-2}$ Hz
maybe the indication that the significant decline in the wave power for low frequencies could be observable with higher spectral resolution.
  
\section{Discussion and conclusions}

As is conventionally accepted, the physical grounds
underlying the appearance of the fast and slow winds are substantially
different. The fast wind originates in the unipolar magnetic configurations
of coronal holes with open field line topology, while most of the indirect
measurements of the active region evolution during the solar cycle, like
detections of the helium abundance in the solar wind, 
indicate that multiple processes for slow wind heating and acceleration
should operate in the different proportions at different phases of the solar
cycle \citep{kasperapj07}. There is the process relate{d to the
heliographic current sheet and the streamer belt
\citep{einaudiasr00,einaudiapj01}, which at the equatorial region is most active at
solar minimum \citep{kasperapj07}. With the uptrend of the activity cycle the significance of the streamer belt
contribution decreases gradually reaching some smaller but finite rates at
maximum. At the same time more and more streamer-like structures appear
at high latitudes covering practically the entire solar disk at maximum activity with relatively small-scale structures.
These structures lead to the cut-off frequencies of the wave spectra generated within
those streamers resulting in reduced acceleration and heating of the plasma outflow. This scenario 
is in good agreement with the observations at high latitudes. 

\begin{figure*}
  \includegraphics[scale=0.9]{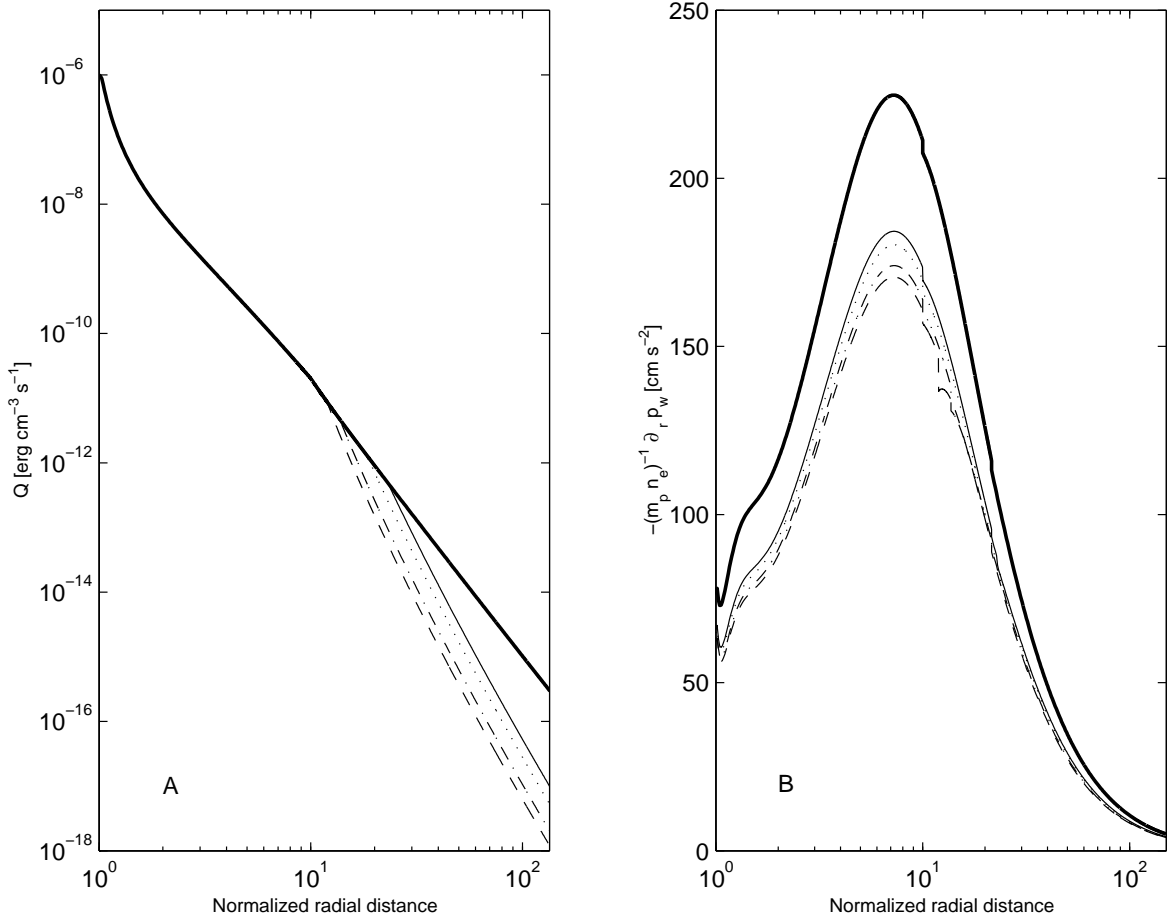}\\
  \caption{Panel A: Calculated heating rates $Q_0$ (reference model, thick solid line), $Q$ (modified 
           heating using the four truncated spectra shown in Fig.~\ref{fig_spectr}. 
           Panel B: Calculated wave pressure gradients for the reference model
           $-(m_p n_e)^{-1} \left(\partial p_{w0} / \partial r \right) $ (thick solid line) and for the truncated spectra
           $-(m_p n_e)^{-1} \left(\partial p_w / \partial r \right)$. The linestyles correspond to Panel A.
           }\label{fig_Q_new}
\end{figure*}

A detailed numerical validation of the presented paradigm of the wave
spectra formation is beyond the primary scope of the current paper where we
focus on the physical grounds of the underlying the concept. While more
extended numerical studies via direct simulations will be published elsewhere, 
here we demonstrate in general terms a correspondence of the above-stated
scenario with the existing data and modern understanding of the solar wind
acceleration scenarios.

At the solar minimum the sources located in the current sheet and streamer belt dominate and the dynamics of the
process is governed by the ambient low frequency Alfv\'en waves \citep{einaudiapj01} for which the
equatorial region is accessible to some extent because of the more regular spatial
structure of the magnetic field below the cusp. The structure of the magnetic field
resembles a dipole field at that stage. With the change of the magnetic
field topology the streamer zone spreads to very high latitudes and, at the same time, 
new open field lines appear which are anchored in the active regions and are fed from the latter 
with the modified spectrum of Alfv\'en wave disturbances \citep{liewersph04,kasperapj07}.
Therefore, the excited wave
spectrum, following the spatial scales of the modified sources, becomes
truncated from the low frequency side. Besides, with the increasing number of
these new active magnetic structures, the rate of confinement of the
plasma at low heliospheric distances leading to a significant modification of
the density profile may also grow. 

In fact, the scenario developed in this paper is an attempt to formulate a mathematical
scheme for a modeling of the slow solar wind throughout the entire latitudinal
domain and throughout the solar activity cycle. Of course, our considerations are based
on substantial assumptions regarding the
fine structure of temporal and latitudinal variation of physical quantities.
More rigorous astrophysical modeling is needed to explore further
specifics of the Alfv\'en wave frequency domain in the solar atmosphere that should
comprise, e.g., the north-south asymmetry of the sunspot distribution,
deviation  from normal latitudinal distributions, or non-equilibrium components
of the observed statistics.
The results can have far reaching impact on the understanding of
the solar and, in general, stellar wind origin. This aspect is also of interest 
in the stellar wind context in connection with the currently emerging
field of stellar wind dynamics and interaction with extrasolar planets.

\begin{acknowledgments}
The results were obtained within the framework of the European Union Research Training Network Solaire
(MTRN-CT-2006-035484). Financial support through the Research Unit 1048 (projects FI 706/8-1/2) funded by the
Deutsche Forschungsgemeinschaft (DFG) is also acknowledged. The work has also been financially suppoerted within the framework
of the FP7-PEOPLE-2010-IRSES project no. 269299 - SOLSPANET.
The authors are thankful to Jens Kleimann, Stefaan Poedts
and Carla Jacobs for valuable discussions. We are grateful to the referee for very constructive comments, which led
to a significant improvement of the manuscript content.
\end{acknowledgments}


\begin{thebibliography}{}
\bibitem[Bamert et al.(2008)]{bamertapjl08}
Bamert, K., Kallenbach, R., le Roux, J. A., Hilchenbach, M., Smith, C. W., and Wurz, P., (2008), ApJ,
{\bf 675}, L45
\bibitem[Chen et al.(2010)]{chenprl10}
Chen, C. H. K., Horbury, T. S., Schekochihin, A. A., Wicks, R. T., Alexandrova, O., and
Mitchell, J., (2010), Phys. Rev. Lett. {\bf 104}, 255002 
\bibitem[Cirtain et~al.(2007)]{Cirtain-etal-2007}
Cirtain J.~W.\ et al., (2007),
Science, {\bf 318}, 1580
\bibitem[dePontieu et~al.(2007)]{dePontieu-etal-2007}
De Pontieu, B.\ et al., (2007),
Science, {\bf 318}, 1574
\bibitem[Einaudi et al.(2000)]{einaudiasr00}
Einaudi G., Boncinelli, P., Dahlburg, R. B., and Karpen, J. T., Advances in Space Research, (2000), {\bf 25},
1931 
\bibitem[Einaudi et al.(2001)]{einaudiapj01}
Einaudi, G., Chibbaro, S., Dahlburg, R. B., and Velli, M., (2001), ApJ, {\bf 547}, 1167
\bibitem[Elfimov et al.(2004)]{elfimovapj04}
Elfimov, A. G.,  Galv\~ao, R. M. O., Jatenco-Pereiredra, V., and  Opher, R., (2004), ApJ, {\bf 600}, 292
\bibitem[van der Holst et al.(2007)]{holstapj07}
van der Holst, B., Jacobs, C., and Poedts, S., (2007), ApJ {\bf 671}, L77
\bibitem[Hu et al.(1999)]{hujgr99}
Hu, Y. Q., Habbal, S. R., and Li, X., (1999), J. Geophys. Res., {\bf 104}, 24819
\bibitem[Kasper et al.(2008)]{kasperprl08}
Kasper, J. C., Lazarus, A. J., and Gary, S. P., (2008), Phys. Rev. Lett. {\bf 101}, 261103
\bibitem[Kasper et al.(2007)]{kasperapj07}
Kasper, J. C., Stevens, M. L., Lazarus, A. J., Steinberg, J. T., and Ogilvie, K. W., (2007), ApJ,
{\bf 660}, 901
\bibitem[Kleimann et al.(2009)]{kleinmanannge09}
Kleimann, J., Kopp, A., Fichtner, H., and Grauer, R., (2009), Annales Geophysicae, {\bf 27}, 989
\bibitem[Liewer et al.(2004)]{liewersph04}
Liewer, P. C., Neugebauer, M., and Zurbuchen, T., (2004), Solar Phys., {\bf 223}, 209
\bibitem[Luo \& Wu(2010)]{luoapjl10}
Luo, Q. Y. and Wu, D. J., (2010), ApJ, {\bf 714}, L138
\bibitem[Marino et al.(2008)]{marinoapjl08}
Marino, R., Sorriso-Valvo, L., Carbone, V., Noullez, A., Bruno, R., and Bavassano, B., (2008), ApJ {\bf 677},
L71
\bibitem[Narain \& Agarwal(1994)]{narainrev94}
Narain, U. and Agarwal, P., (1994), Bulletin of the Astronomical Society of India, {\bf 22}, 111
\bibitem[Oughton \& Matthaeus(2005)]{oughtmatt05}
Oughton, S., and Matthaeus, W. H., (2005), Nonlin. Processes Geophys., {\bf 12},
299
\bibitem[Richardson et al.(2001)]{richardson01}
Richardson, J. D., Wang C., and Paularena, K.I., (2001), Adv. Space. Res, {\bf 27}, 471 
\bibitem[Sahraoui et al.(2009)]{sahraouiprl09}
Sahraoui, F., Goldstein, M. L., Robert, P., and  Khotyaintsev, Y. V., (2009), Phys. Rev. Lett.
{\bf 102}, 231102 .
\bibitem[Shergelashvili et al.(2006)]{shergself06}
Shergelashvili, B. M., Poedts, S., and Pataraya, A. D., (2006), ApJ, {\bf 642}, L73
\bibitem[Tu et al.(1984)]{tujgr84}
Tu, C., Pu, Z., and  Wei, F., (1984), J. Geophys. Res. {\bf 89}, 9695 .
\bibitem[Tu \& Marsch(1996)]{tumarsch96}
Tu, C. and Marsch, E., (1996), Space Sci. Rev. {\bf 77}, 372 
\bibitem[Ulmschneider et al.(2001)]{ulmshneiderapj01}
Ulmschneider, P., Fawzy, D., Musielak, Z. E., and  Stepie\'n, K., (2001), ApJ {\bf 559}, L167 
\bibitem[Vainio et al.(2003)]{vainioaa03}
Vainio, R., Laitinen, T., and Fichtner, H., (2003), A\&A {\bf 407}, 713
\end{thebibliography}
\end{document}